\documentclass[11pt,twocolumn]{article}
\usepackage{amsmath,amsfonts,xcolor,endnotes,abstract}

\usepackage{graphicx}
    \setkeys{Gin}{width=.75\textwidth}

\let\footnote\endnote
\usepackage{hyperref}
\usepackage[paperwidth=8.5in, paperheight=11in]{geometry}
\begin{document}

\title{\textbf{Designing and Building a Three-dimensional Projective Scanner for Smartphones }}
\author{Papachristou Marios \\ 
\small{P2P Lab, Ioannina \& P2P Foundation} 
}
\date{}

\twocolumn [ 
\maketitle 
\begin{abstract}
One of the frustrating things in the digital fabrication era is that its media are neither affordable nor easily accessible and usable. Three-dimensional (3D) fabrication media (DFM) such as 3D Printers and 3D Scanners have experienced an upsurge in popularity, while the latter remain expensive and hard to function. With this paper, we aim to present you the RhoScanner Project - a an affordable and efficient Three-dimensional Projective Scanner for Smart-phones, hence shedding light on the extended capabilities of digital fabrication media on popular use.
\end{abstract}
]

\section{Introduction}
Nowadays, the need for digital fabrication  (DF) devices is constantly rising. The emergence of such techniques and technologies on a popular basis, like three-dimensional (3D) printing and scanning has been gaining ground the recent years, thence it has been a common problem that even if the costs of 3D Printers are constantly reducing, the costs of 3D Scanners have not changed so radically according to \cite{straub14}. Thereby, an austere structure whose goal is to minimize materialization costs, empowered with the extended capabilities of free and open-source software targeted to mobile devices is a model that perfectly fits the current situation. 

Having this in mind, we decided to reverse-engineer a 3D scanner and create an opensource project under the name \emph{RhoScanner}\endnote{\url{http://rhoscanner-team.github.io}. Accessed December 2014} which consists of the following tools: a) \emph{RhoScanner} (The Core Scanner Application), b) \emph{RhoScanner Viewer} (A viewer for 3D files that covers the mere basics of OBJ file viewing), a \emph{PCD Plotter} (A simple scatter plotter for plotting point data using Matplotlib) and a simple \emph{OBJ Viewer} for viewing OBJ files.

The project was developed in Python which is an open-source object-oriented multi-platform programming language, with a large number of libraries (e.g. for numerical analysis, computer vision etc.) suitable to our needs. We chose to use Python because it offered us the desirable results due to its portability and elegant programming style, whereas the ability to improve the performance of the algorithms is full-fledged with tools like Cython. We have also included Kivy, a cross-platform open-source library for Python for multi-touch application development with a Natural User Interface (NUI).   

Finally, this paper aims to provide you with a basic guidance to the project as it contains the basic mathematical and algorithmic concepts behind the ensemble. the In Section \ref{sec:outline}, we present a brief outline of the project as well as the construction/implementation timeline. In Section \ref{sec:math}, we dive into the Mathematical Background and describe the essential transformations in theory which are being used to generate the pointclouds (PCDs). Section \ref{sec:cs} is targeted to the algorithmic implementation of the project. Besides, it examines the optimizations being done - or the possible ones - to the algorithm for better performance. We, then, depict the Hardware parts in Section \ref{sec:hwd} and present our Results in Section 5. Finally, Section \ref{sec:conc} is the place where you can find the conclusions of the project. 

\section{An Outline of the Project}
\label{sec:outline}

\subsection{Description}

As described in the P2P Foundation Wiki\endnote{\url{http://p2pfoundation.net/RhoScanner}. Accessed December 2014}:

\begin{quote}
\textit{``RhoScanner is a projective mobile fully-functional three-dimensional scanner that costs only 12USD. RhoScanner is a completely reverse-engineered project being maintained by (...) under \emph{http://rhoscanner-team.github.io}. It is being developed purely in Python and licensed under the GNU General Public License v3. Collaboration requests are more than welcome.'' }

\end{quote}

generically states that RhoScanner is \emph{projective}, since it uses a line LASER (LL) to accomplish curve projection. The LL is targeted to the scannable object and the deformation of the projected line is accurately measured (see Section \ref{sec:math}) Additionally, it is \emph{mobile}; designed and built to run on a mobile phone (an Android Smartphone), therefore enriching the mobility of DF media as well as \emph{affordable}, costing only 12USD. The scanner was also \emph{reverse-engineered}, based on the observation of such machines, meaning that it might be partly a novel approach to previous work on 3D Scanners, \emph{modular} with both hardware and software extended modification capabilities due to the portability of Python and the tools it uses (Python can run both on Android and iOS devices), while it remains purely free and opensource software accesible and modifiable by everyone considering that it is \emph{licensed under the GNU General Public License v3} with having its source code released via Github CVS.

\subsection{Timeline}

The project was initiated at March, 2014. At the beginning, the core libraries were written. These libraries included: a) \emph{a mini computer vision library} able to handle basic operations on images such as shifting colour modes between RGB, HSV, HLS and YIQ and thresholding and b) \emph{a module for handling mathematical computations} as explained in Section 2. Thereafter, the core libraries were ready and we proceeded developing a graphical user interface and the main frame, where RhoScanner would operate (see Sections \ref{sec:cs}, \ref{sec:hwd}).   

\section{Mathematical Background}
\label{sec:math}

\subsection{Image Preparation}

Any image captured by the camera is represented as an $m \times n$ matrix. Therefore, let $$ I = \begin{bmatrix}
 p_{11} & p_{12} & \cdots & p_{1n} \\
 p_{21} & p_{22} & \cdots & p_{2n} \\
 \vdots & \vdots & \ddots & \vdots \\
 p_{m1} & p_{m2} & \cdots & p_{mn}
 \end{bmatrix}$$

be the matrix of the image and $p_{ij}$ a random pixel at $(i,j),   \quad i \leq m \;, j \leq n $. The characteristics of this pixel are $$p_{ij} = (R_{ij},G_{ij},B_{ij})$$ where $R_{ij},G_{ij},B_{ij}$ are the values of the red, green and blue channels respectively. 

Furthermore, let $T(v,\alpha)$ be the threshold function applied on a pixel so that:

$$T(v,\alpha) = \bigg\{ \begin{matrix}
 255 & , & v \geq \alpha  \\
 0 &  , & v < \alpha 

\end{matrix}
 $$ 

For each image frame that is captured the algorithm filters the red channel of the image and applies a threshold such that the new image  $I_T = [p_{T \; ij}]$ such that $$V(p_T) = T( (V(p)), \alpha_0) \; \forall p \in \mathrm R (I), \; \alpha_0 \in (0,255) $$

where $V(p)$ represents the value of the pixel of the red channel of $I$ (or $\mathrm R (I)$)
\subsection{Curve Transformations} 

Behind the whole process of development and implementation, a mathematical background was set in order to study the curves and the operations needed for getting satisfying results. We studied curve geometric transformations which could generate the appropriate results for obtaining tuples with three elements (point clouds)

\begin{figure}
  \centering
    \includegraphics[]{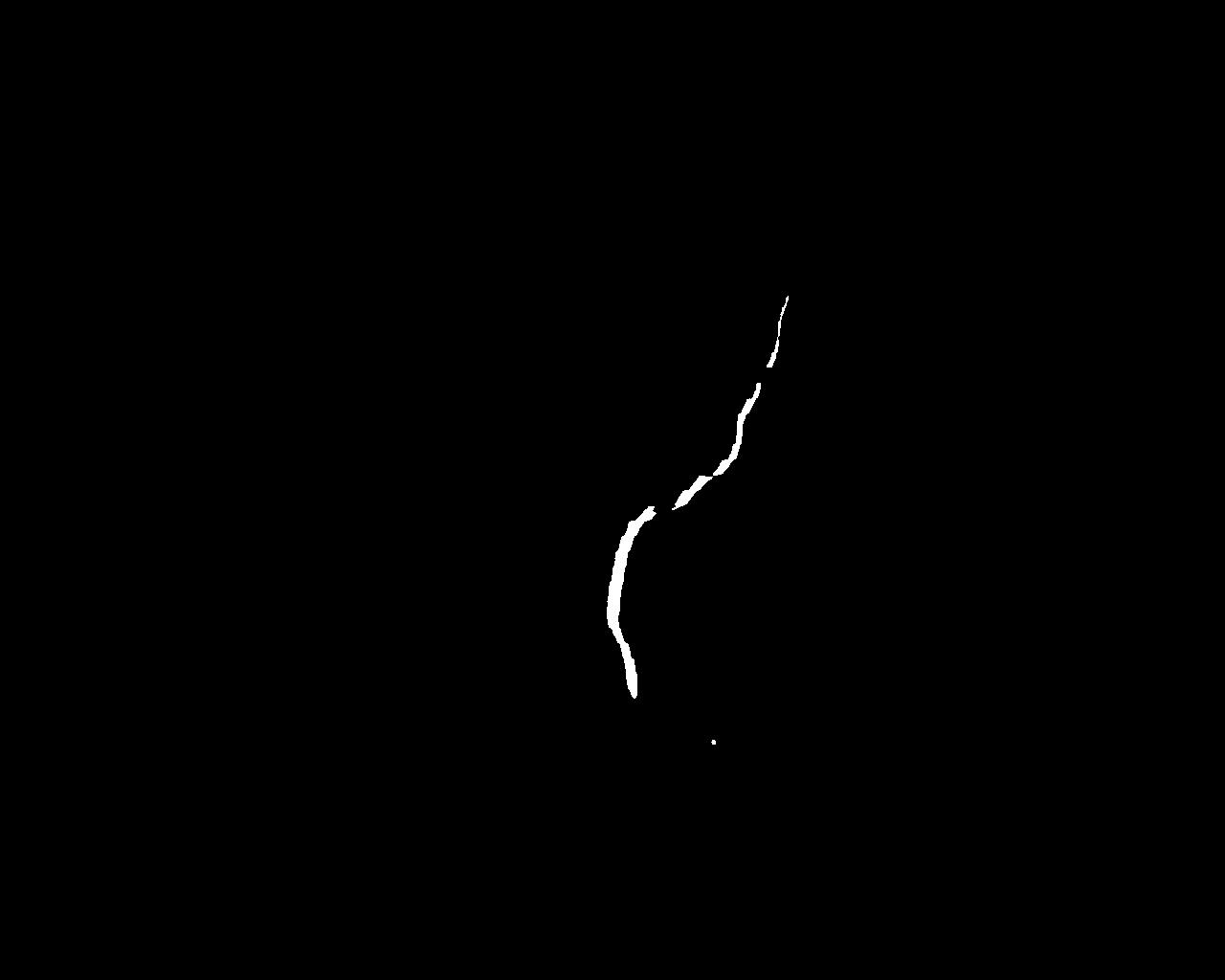}
  \caption{An arbitrary initial poinset $\mathbb{S}^2$.}
    \label{curve1}
\end{figure}

The aim of three dimensional scanning is to obtain a three dimensional point cloud (PCD), namely a smooth curve that contains all the essential data in order to be converted to a mesh through a triangulation. So, let

$$ \mathbb S^2 = \bigcup^b_{i=a} \{ P_i \} $$

be the set of points obtained by the camera after applying the threshold function such that $S^2 \subseteq \bigcup_{i=1}^n \bigcup_{j=1}^m \{ p_{T \; ij} \}$. The following image is an actual representation of what is obtained by the camera where the white pixels represent the red points of the point-set (Figure \ref{curve1}).

\begin{figure}
  \centering
    \includegraphics[]{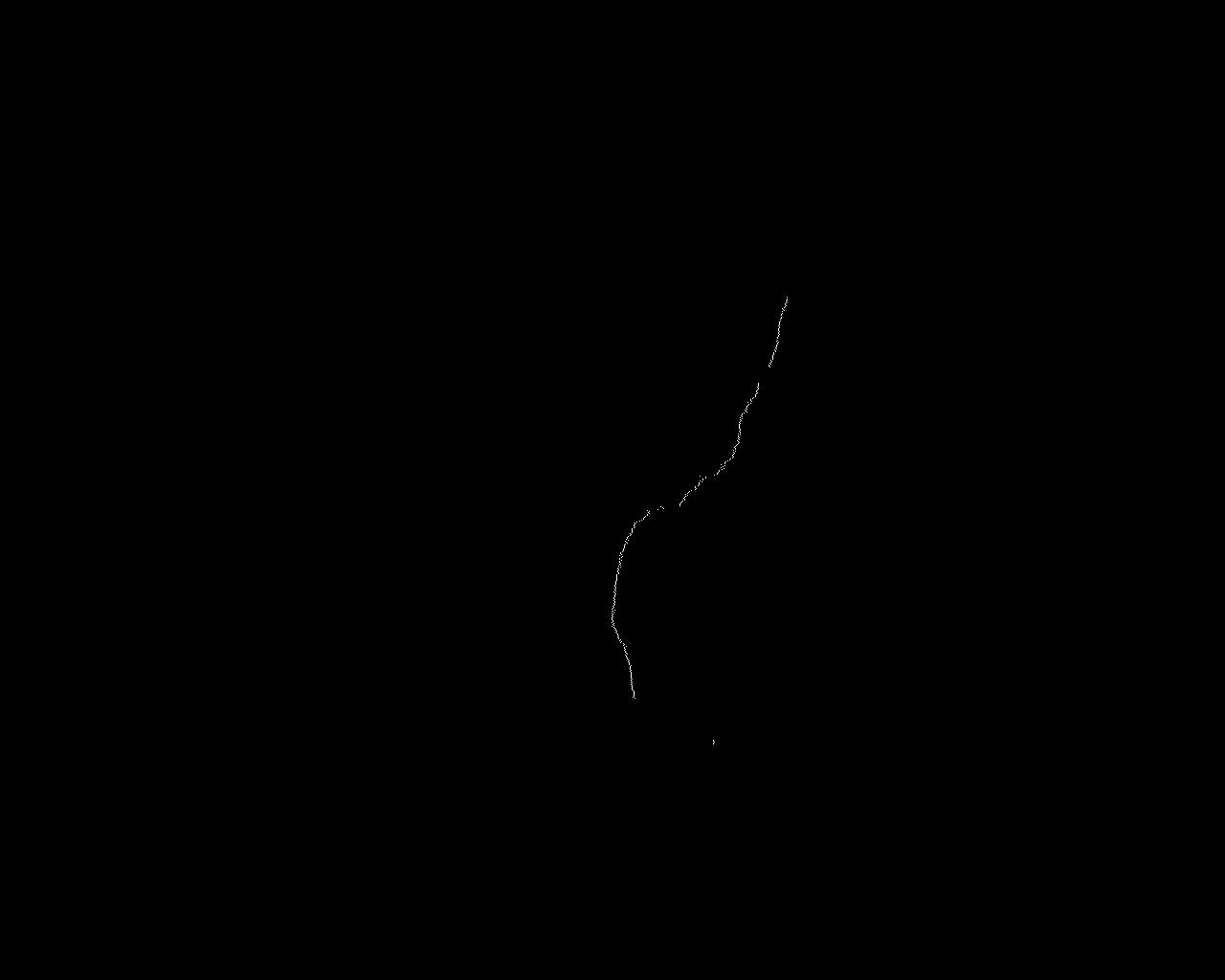}
  \caption{The resulting curve $x=x(y)$ after $T_1$}
    \label{curve2}
\end{figure}

The first transformation is used to convert this set to a map (curve) $C^2:x \to x(y)$ in the two-dimensional (2D) space and a point $M$ on the curve. As the demanded result is derived by observations\footnote{data} close to one another, the best way of smoothing the curve was to use the arithmetic mean of the $x$ values while $y$ is held constant. Note that the XY plane is similar to the XY plane of a normal image. Therefore, the transformation is the following: $$ T_1 :  \mathbb{S}^2 \to C^2 $$

according to which

$$M(x_i,y_i) \to M' \bigg( \frac{1}{m} \sum^m_{j=k} x_{ij} , y_i \bigg) \equiv M'(\bar x_i,y_i)$$

resulting to the smoothed curve of Figure \ref{curve2}.

\begin{figure}
  \centering
    \includegraphics[]{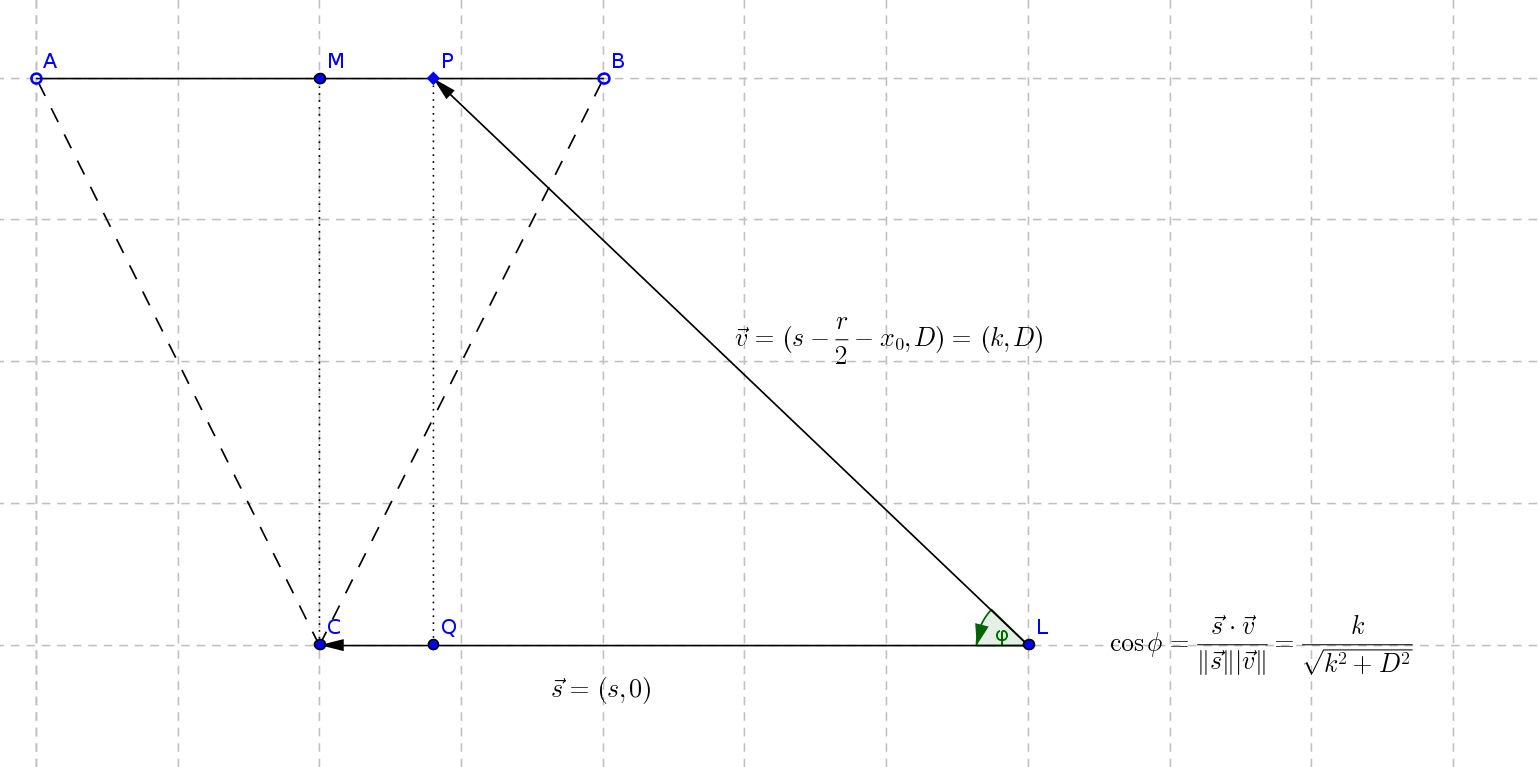}
  \caption{The angular calculations}
    \label{curve3}
\end{figure}

Moreover, we used a rotary transformation $\mathrm{Rot(\theta)}$ in order to convert the the 2D curve to a 3D one, according to which the points are being rotated by a specific angle. To do this we introduced two vectors $\vec s = (s,0)$ and $\vec v = \big ( s-\frac{r}{2}-x_0,D \big ) = (k,D)$ (as seen in Figure \ref{curve3}). Hence, the angle is obtained via the arc cosine function of $$\cos{\phi} = \frac{\vec s \cdot \vec v}{\|\vec v\|\|\vec s\|}$$

implying that $\theta$ is equal to \footnote{$\phi+\theta=90^\circ$}

$$\theta = 90^{\circ} - \arccos{ \bigg (\frac{k}{\sqrt{k^2 + D^2}} \bigg ) }  $$

\begin{itemize}
\item $s$ is the distance between the laser and the camera ($s>0$)
\item $D$ is the distance between the flat surface and the camera
\item $r$ is the Y range of the camera
\item $x_0$ is the minimum X coordinate
\end{itemize}

which means that $M(x,y) \xrightarrow {\mathrm{Rot} (\theta)}  M'(x \cos \theta, y \sin \theta)$

The final transformation $$T_2: C^2 \to C^3 $$ is the one that results in a three-dimensional curve. In more detail, it does the following operations for every point M of the two-dimensional map (curve)\footnote{$C^3:z=f(x,y)$} 

$$M (x,y) \to M'(x_0',y,D- \Delta x) $$

where $\Delta x = x - x_0'$ and $x_0'$ is the minimum $x$ coordinate after $\mathrm{Rot}(\theta)$ resulting in returning three dimensional points $P(x,y,z)$ which are afterwards saved as raw data in an ASCII file.

\section{Algorithm Implementation and Optimization Techniques and Possibilities}
\label{sec:cs}

\subsection{Postulations on using \texttt{cython} to optimize the libraries}


As stated in \cite{cython11}, \emph{Cython} is an extension to the programming language Python that allows for explicit type declarations as well as porting Python code to native C (or C++) code.Therefore compiled Cython Code, which stands for the code written with Cython and compiled with a C/C++ compiler, behaves and performs better than Python Code by using the power of the pre-existing C and FORTRAN code. For example, large numerical loops finish faster due to native type declaration. [again \cite{cython11}]

It is also essential to understand that the RhoScanner application does thousands of such computations when an object is being scanned. So, the need for fast computations in a relatively short(er) time is obvious given that few smart-phones have high computation power nowadays. Hence, the materialization of this project using the Cython language will arise a computation upsurge. Our work on porting Cython code to compiled C code consists of a small python script\endnote{\url{http://goo.gl/EWhxb8}. The \texttt{android-cythonizer} script. Accessed December 2014} that converts Cython code to C code and uses \texttt{arm-linux-androideabi-gcc} or \texttt{arm-linux-gnueabi-gcc} to compile the code for the ARM architecture. Thenceforth, the shared libraries that were generated can then be imported either in an interpreted environment, like \emph{QPython}, or pre-compiled environments, built with \emph{python-for-android}.

\subsection{Developing graphical front-ends with \texttt{kivy}}

The Graphical User Interface (GUI) of the Rhoscanner Application is a button-layered interface developed using the Kivy Language - an XML-like language - for widget development.
Also, the OBJ Viewer is also based on the Kivy Language which is responsible for displaying the scene.

\section{Hardware \& Designs}
\label{sec:hwd}

\begin{figure}
  \centering
    \includegraphics[]{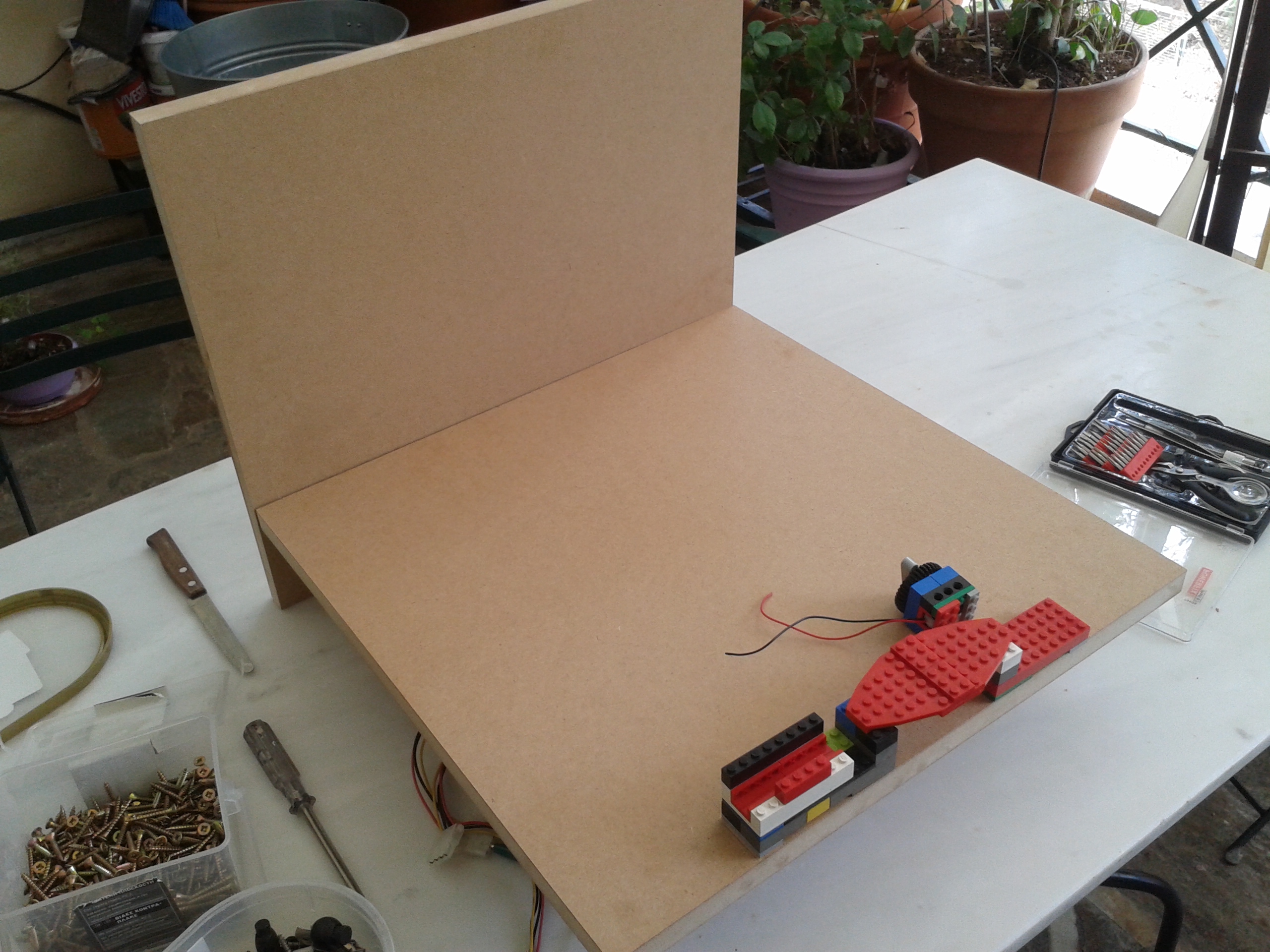}
  \caption{The frame with the LEGO-built holder mounted on it}
    \label{figure4}
\end{figure}

\begin{figure}
  \centering
    \includegraphics[]{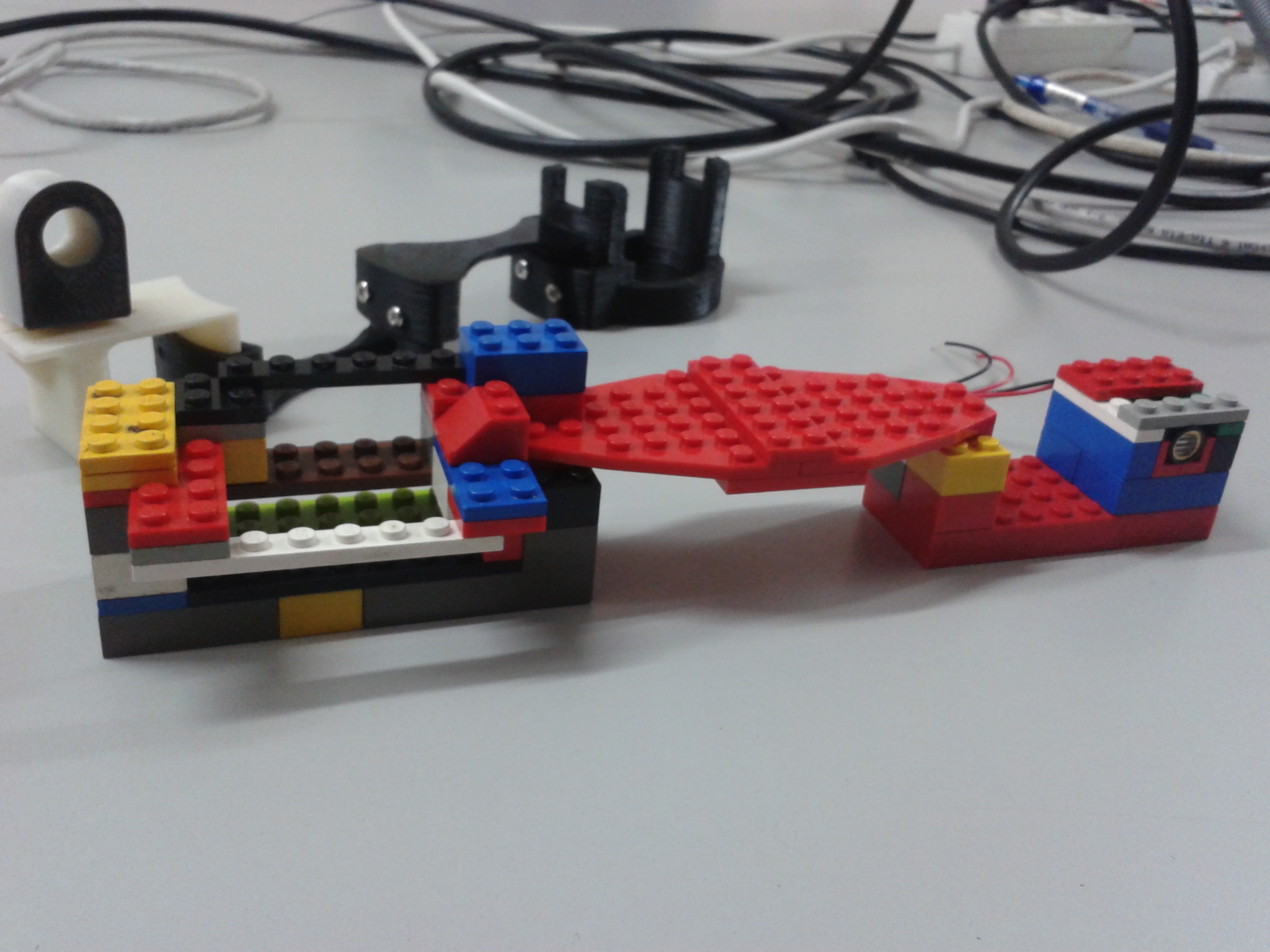}
  \caption{Close-up photograph of the LEGO-built holder which is as large as a MakerScanner 3D-printedframe}
    \label{figure5}
\end{figure}

We assembled a base from MDF by putting two MDF plates ($40 \times 40 \mathrm{cm}$, $40 \times 45 \mathrm{cm}$) under a right angle. Then the smart-phone-laser holder was built with LEGO and was mounted on the MDF system (Figures \ref{figure4},\ref{figure5}). The LL and smart-phone were mounted onto the holder  Additionally, a 3D model of the smart-phone-laser holder was designed and uploaded to \emph{Thingiverse}\endnote{\url{http://www.thingiverse.com/thing:315116}. Three-dimensional-printable case derived from MakerScanner. Accessed December 2014  }; available for 3D printing. 
The tests were run on a 1 GHz Cortex-A5, 645MB RAM smart-phone running Android 4.0.4. Optionally, an Arduino/RaspberryPi could be responsible for shifting the LL.

\section{Results and Discussion}
\label{sec:results}

\subsection{System Operations for artifact scanning}

\begin{figure}
 \centering
    \includegraphics[]{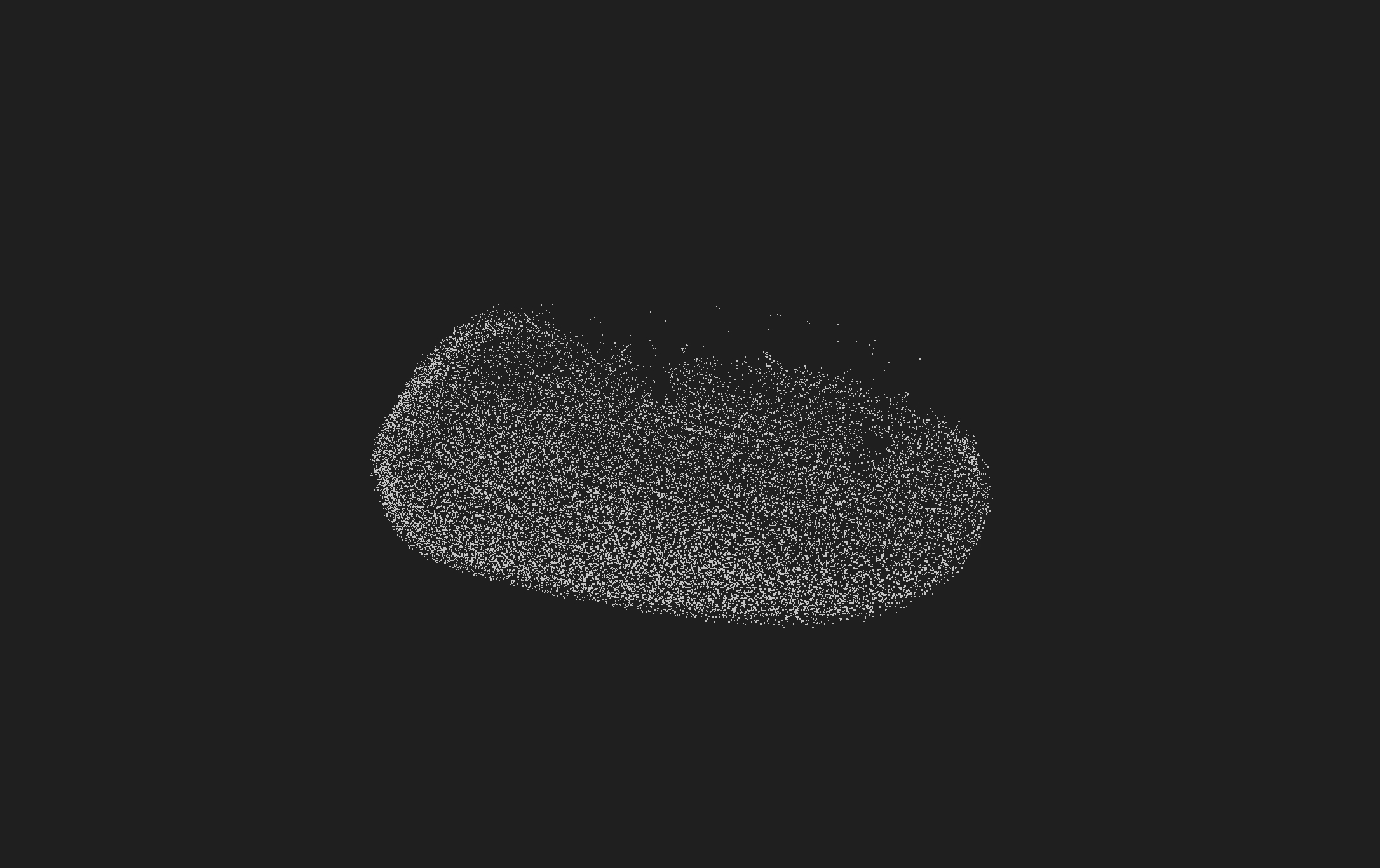}
  \caption{Some scanned artifact. Post-editing stage. (merging, noise reduction etc.)}
    \label{artifact}
\end{figure}

In this part of the paper, we attempt to present the system operations behind scanning artifacts as well as the artifacts themselves, from a holistic perspective. First, the artifact is placed on the main frame with the LL pointing at one of its sides. Then, the scan begins in a dark room for best results (capturing the red channels). After each frame-scan, the LL is slowly shifted using an appropriate mechanism (e.g. an Arduino with a 5V servo/stepper motor or even carefully by hand) to the next scanning position. The process is repeated until the LL points the other side of the object. The time depends as well heavily on the algorithm itself and the resolution $\Delta \theta$ of the shifting mechanism which is the change in the angle of the LL for every frame. Our approach, was able obtain two point clouds (the scan is repeated twice for each side -front and rear - of the object) of the same object and, after a process of noise reduction and point cloud merging which was done, for the time being, by hand, we are able to acquire the final result, namely the complete artifact (see Figure \ref{artifact}).

\section{Conclusions and Future Work}
\label{sec:conc}

With this paper, we have managed to present the design, mathematical background and development process of a low-cost three-dimensional scanner being developed elaborating open-source standards and rapid prototyping in the process of materialization and distribution via the GNU General Public License v3 License. This project has synopsised existing work related to three dimensional scanners and the rapid manufacturing techniques in means of simplicity and giving ground to Commons-based Peer Production \cite{kostpapa14} and the Maker Movement. Our paper has described the physical structure, the algorithmic procedure as well as postulations on using software like Cython to improve performance on Python-based applications that run on ARM devices. Future work will include writing better performing algorithms and feature enrichment such as porting triangulation algorithms to compatible out-of-the box executable code using computational geometry software (Computational Geometry Algorithms Library also known as CGAL or PCL; prefferably in C/C++ code) being bridged to application via Cython/\texttt{ctypes}. The characterization of the performance of the scanner under various configurations will also be taken into consideration.

All in all, a modular scanner that uses completely free and open source software and open hardware costing approximately 12USD that anyone can build - preferably in a local Makerspace like a hackerspace where three dimensional printing is broadly available - enriches, undoubtedly, the Internet of Things whose aim is to support rapid manufacturing of new products, dynamic response to product demands, and real-time optimization of manufacturing production and supply chain networks, by networking machinery, sensors and control systems together, as referred in \cite{IoT13}. Another possible use, besides this, is to educate people in graphics, mathematics and computer science courses.

\section*{Acknowledgements}

The author would like to thank his fellows at P2P Lab and P2P Foundation for their contributions in mentoring his work and the people who contributed in reviewing this paper's preprints.


\begingroup
\theendnotes
\endgroup

\section*{About the author}

\paragraph{Marios Papachristou} Marios was born in Athens (GR) in 1997. He is currently a high school student. He has been contributing in open-source and open-hardware projects since 2010, being an ardent open-source fan. He has worked on developing digital fabrication media as well as bioinformatics applications using computer vision. He published his first paper at the age of 16, while he remains active by contributing to many projects.

\onecolumn

\newpage

\section*{Appendix}

\end{document}